\documentclass[namedreferences]{SolarPhysics}
\usepackage[optionalrh,solaenum]{spr-sola-addons} 
\usepackage{graphicx}        
\usepackage{color}           
\usepackage{url}             
\newcommand{\etal}{{\it et al.}}

\newcommand{\aap}{    {\it Astron. Astrophys.}}

\newcommand{\apj}{    {\it Astrophys. J.}}
\newcommand{\apjl}{   {\it Astrophys. J. Let.}}
\newcommand{\apjs}{   {\it Astrophys. J. Suppl. Ser.}}

\newcommand{\araa}{   {\it Ann. Rev. of Astron. Astrophys.}}

\newcommand{\grl}{    {\it Geophys. Res. Lett.}}

\newcommand{\solphys}{{\it Solar Phys.}}

\begin{document}

\begin{article}

\begin{opening}

\title{Recent Developments of NEMO: Detection of Solar Eruptions Characteristics \\ 
}

\author{O.~\surname{Podladchikova}$^{1}$\sep
        A.~\surname{Vuets}$^{1}$\sep
        P.~\surname{Leontiev}$^{1}$\sep
        R. A. M.~\surname{Van der Linden}$^{1}$
       }
\runningauthor{O.~Podladchikova \etal}
\runningtitle{NEMO Detection of Solar Eruptions}

   \institute{$^{1}$ The Solar-Terrestrial Centre of Excellence, Royal Observatory of Belgium, Ringlaan-3, 1180--Brussels, Belgium
                     email: \url{Elena.Podladchikova@stce.be}\\
              $^{2}$ Institute of Applied System Analysis, National Polytechnic University of Ukraine, Pobeda 37, 03056-Kiev, Ukraine\\
             }

\begin{abstract}
The recent developments in space instrumentation for solar observations and telemetry have caused the
necessity of advanced pattern recognition tools for the different classes of solar events. The Extreme ultraviolet Imaging Telescope (EIT) of solar corona on-board SOHO spacecraft has uncovered a new class of eruptive events which are often identified as signatures of Coronal Mass Ejection (CME) initiations on solar disk. It is evident that a crucial task is the development of an automatic detection tool of CMEs precursors.

The Novel EIT wave Machine Observing (NEMO) ({\color{blue}\url{http://sidc.be/nemo}}) code is an
operational tool that detects automatically solar eruptions using EIT image sequences. NEMO applies techniques based on the general statistical properties of the underlying physical mechanisms of eruptive events
on the solar disc. In this work, the most recent updates of NEMO code - that have resulted to the increase of the recognition efficiency of solar eruptions linked to CMEs - are presented. These updates provide calculations of the surface of the dimming region, implement
novel clustering technique for the dimmings and set new criteria to flag the eruptive dimmings
based on their complex characteristics. The efficiency of NEMO has been increased significantly resulting to
the extraction of dimmings observed near the solar limb and to the detection of
small-scale events as well. As a consequence, the detection efficiency of CMEs precursors
and the forecasts of CMEs have been drastically improved. Furthermore, the catalogues of solar eruptive events that can be constructed
by NEMO may include larger number of physical parameters associated to the dimming regions.

\end{abstract}
\keywords{Corona $\cdot$ Instrumentation and Data Management, Coronal Mass Ejections $\cdot$ Low Coronal Signatures $\cdot$ Flares }
\end{opening}
\section{Introduction}

{Present knowledge about susceptibility of modern technological systems to space weather related disturbances has lead in a significant rise of the scientific and operational interest of issues associated to the understanding and forecast of space weather events.  In turn, it has resulted in substantial increase of simultaneously operating satellites devoted to space weather effects, and to an exponential growth in the volume of space weather related data. Thus, it is impossible to provide comprehensive analysis of these huge quantity of data without the development of proper algorithms and software tools that are able to identify automatically segments of data relevant to a particular physical process of space weather event of interest.}

{The sudden energy release of the solar corona is the source of powerful electromagnetic disturbances of the near-terrestrial environment. Solar flares and Coronal Mass Ejections (CMEs) can trigger geomagnetic storms which may affect terrestrial communications and the reliability of power systems. The most hazardous space weather events are associated with CMEs \cite{Gopal2004,Gosling1990,Kahler1992,Richardson2001,Zhukov2004a}.
Thus, CMEs and flares are principal objects of detection and cataloguing. However, automatic detection of flares and CMEs detection is not evident,
due to their complex characteristics. A number of flare and CME detection methods in different wavelengths with different instrumental limitation
 have been reviewed by \inlinecite{RobbrechtRew2005}. Today a number of the instruments accessible for space community is able to monitor flares and CME events on the regular basis. Moreover, there are several data sources that can be consulted in real or near-real time for the case of solar flares.
The Solar Flare Automatic Detector of Solar Soft, which is based on recognition of total X-Ray intensity flux bursts
from GOES-2, provides a solar flare catalog which is automatically constructed and is available on LMSAL/NASA website ({\color{blue}\url{http://www.lmsal.com/solarsoft/latest_events}}). A Full catalog of solar flares observed by different
instruments is manually constructed by NOAA Space Weather Center in Boulder/US with information provided to the community
with one day delay ({\color{blue}\url{http://www.swpc.noaa.gov/ftpdir/indices/events}}). The detection of CMEs has been traditionally based on the manual recognition  of features moving radially outwards from the Sun using coronagraph data. The Full CME catalog based on LASCO/SOHO coronograph
observations is manually constructed by N. Gopalswamy group ({\color{blue}\url{http://cdaw.gsfc.nasa.gov/CME_list}}).
A significant progress in the automatic detection and cataloguing of CMEs was achieved recently by the CACTUS software
using LASCO/SOHO and SECCHI/STEREO data ({\color{blue}\url{http://sidc.be/cactus}}) \cite{Robbrecht2004}.
This has led to issuing of near-real-time messages (with a delay of a quarter of a day) alerting the space weather community.}

{The ever growing importance of space weather has led to new requirements on the accuracy of CME detection and forecast.
In situ measurements are not able to address the problem of prediction of solar eruptive events that can be extremely hazardous to space based technological systems. However, solar surface observations may provide such information in advance and substantially increase the forecast time of CME prediction. Furthermore, the increasing processing power of computers and the evolution in image processing and pattern recognition techniques has made possible to develop automatic detection tools for the solar drivers of space weather event.}

{During the last decade the Extreme UltraViolet (EUV) solar imaging from space has revealed a rich diversity of solar disk events,
such as global waves and eruptive dimmings (so-called "EIT wave" phenomenon \cite{Thompson1998}.
EUV Imager Telescope on-board SOHO \cite{boudin} was able to transmit high cadence solar corona EUV images on Earth from the beginning of 1997 till 1 August, 2010 24 hrs a day. EIT waves appear as global bright features propagating in the solar corona followed by intensity dimness (dimming) as seen in Extreme UltraViolet (EUV).
They are usually triggered by solar flares or brusque filament disappearance.
It has been suggested that these waves are among the best indicators of the large-scale reorganization of coronal magnetic fields \cite{Plunkett1998,Biesecker2002,Harra2001}. Various EIT waves were found to precede CMEs in time and space  (i.e. CME signatures) and later got a scientific name of Solar Eruptions. EUV observation of the associated with CME phenomena (e.g., eruptive dimmings and EIT waves) can be processed faster than coronagraphic images (e.g., less images are required for  detection). Thus, observations of eruptive dimmings end EIT wave
phenomena might improve drastically the forecast times of CME warnings.}

{The first catalog of the large scale
EIT waves was manually constructed by \inlinecite{Thompson2009} and accounted for the years 1997--1998.
In contrast to the detection of solar flares and CMEs which is based on the recognition of well distinctive
signatures (intensity bursts or global structures propagating in the plane of sky) the detection
of EIT waves is not trivial since they are characterized by rather low intensity (on the level of general background), they
present large variety in morphology and can last down to tens of minutes.}

{The first proof-of-principle demonstration for the automated detection of EIT waves
using the statistical properties of the eruptive EUV on-disk events and the underlying
physical mechanisms was proposed by \inlinecite{Epod2005a}. The physical properties of eruptive CME on-disk precursors
modify drastically the high order statistical properties of the image sequences.
Based on this proof-of-principle an algorithm was developed in order to detect EIT waves
and eruptive dimmings. The algorithm was successfully tested with the calibrated EIT/SOHO data of 1997--1998 period
in order to extract the events listed in the EIT wave catalogue of \inlinecite{Thompson2009}.
In 2006, the Novel EIT wave Machine Observing (NEMO) ({\color{blue}\url{http://sidc.be/nemo}})
operational tool was developed in order to detect automatically solar eruptions using real-time quick look
EIT images and extract eruptive dimmings with their parameters. One of the particularly important NEMO feature
is the capability to detect among the large family of solar dimmings - which are always present on the EUV solar disk -
those connected to CMEs. NEMO has been built using a series of high level image processing
techniques suitable to extract eruptive features from the EUV solar disk under complex solar conditions.

The operation of NEMO allowed the automatic construction of a catalogue with the Eruptive Dimmings and EIT waves
as detected in the image sequences of SOHO for the period 1997--2010. The majority of the events listed in the
catalogue were identified as CME precursors (using CME catalogs). In contrast to the detection methods based on coronograph data, NEMO could detect even the precursors of faint halo CMEs as well.}

{EUV solar imagers are currently operating on-board Proba-2, STEREO (Solar TErrestrial RElations Observatory) and SDO (Solar Dynamic Observatory) missions while two more imagers are under construction for future space experiments. A project for automatic recognition of Eruptive Dimmings for post-processed SDO data was recently initiated by NASA \cite{Martens2011} using detection principles similar to that of NEMO \cite{Attrill2010}.
Similar principles have been also used for automated flare detection based on EUV images by \inlinecite{Patoul2008} while
\inlinecite{Slemzin2005} demonstrated a principle to detect eruptive dimmings that are widely distributed in the visible
EUV solar corona.}

{It is evident that NEMO may further contribute to the development of detection schemes of eruptive events.
In this work, the recent updates of NEMO code are presented resulting to increase of the recognition efficiency
of the solar eruptions linked to CMEs. In particular, these updates provide: (1) the direct calculation of the
the surface of the dimming region in terms of physical variables (square kilometers) since
EIT images are in fact projection of the solar sphere;
(2) the optimization of a clustering technique for the dimmings;
(3) new criteria to flag the eruptive dimmings, based on their complex characteristics (area and intensity of dimmings).
The basic scheme of NEMO algorithm is described in the next section and the most recent developments in section 3.
Some examples are presented in Section 4 and a conclusive summary in the last section.}

\section{The NEMO algorithm}

%
%

\begin{figure}    
   \centerline{\hspace*{0.01\textwidth}
               \includegraphics[width=0.40\textwidth,clip=]{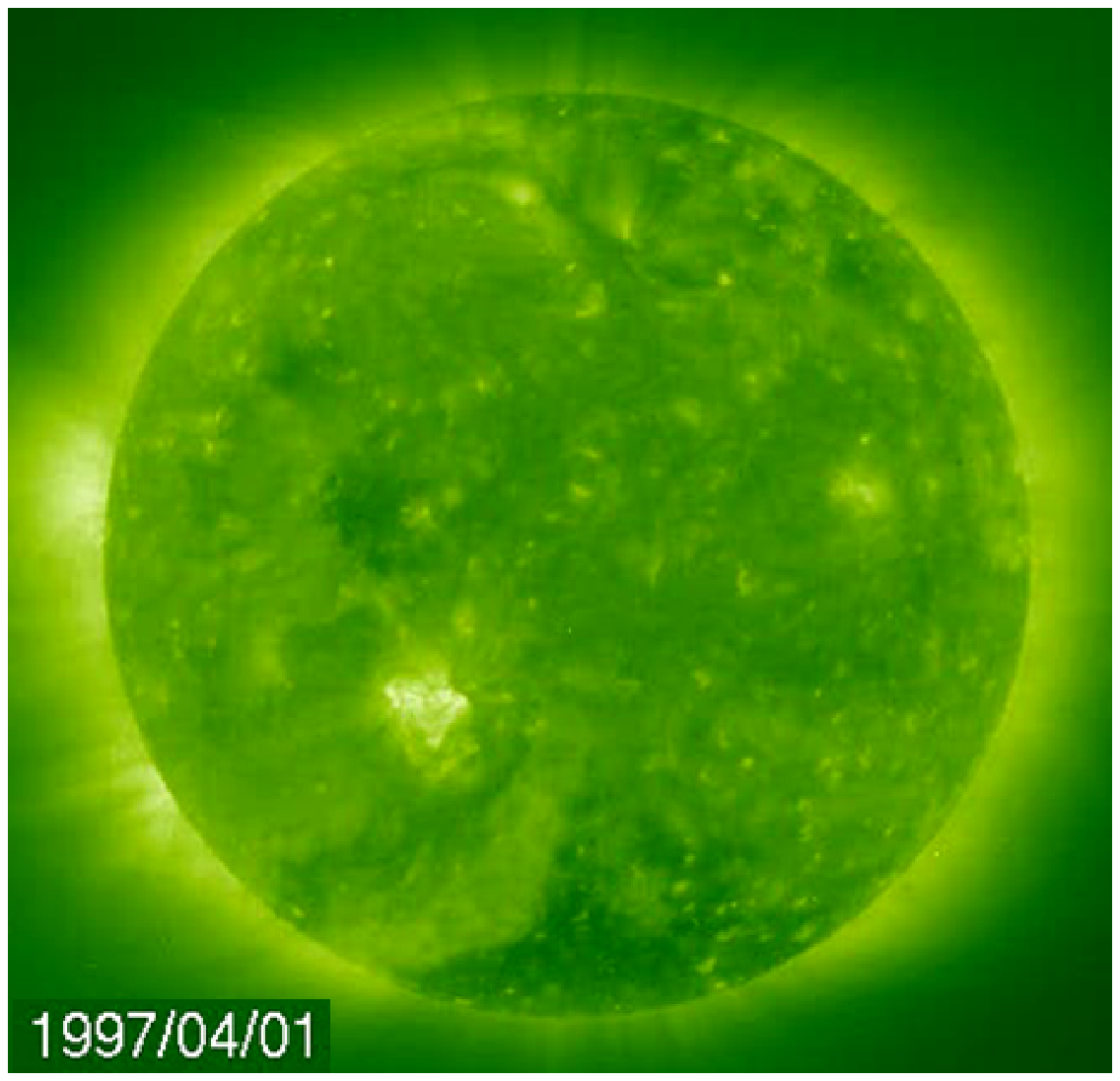}
               \hspace*{-0.014\textwidth}
               \includegraphics[width=0.525\textwidth,clip=]{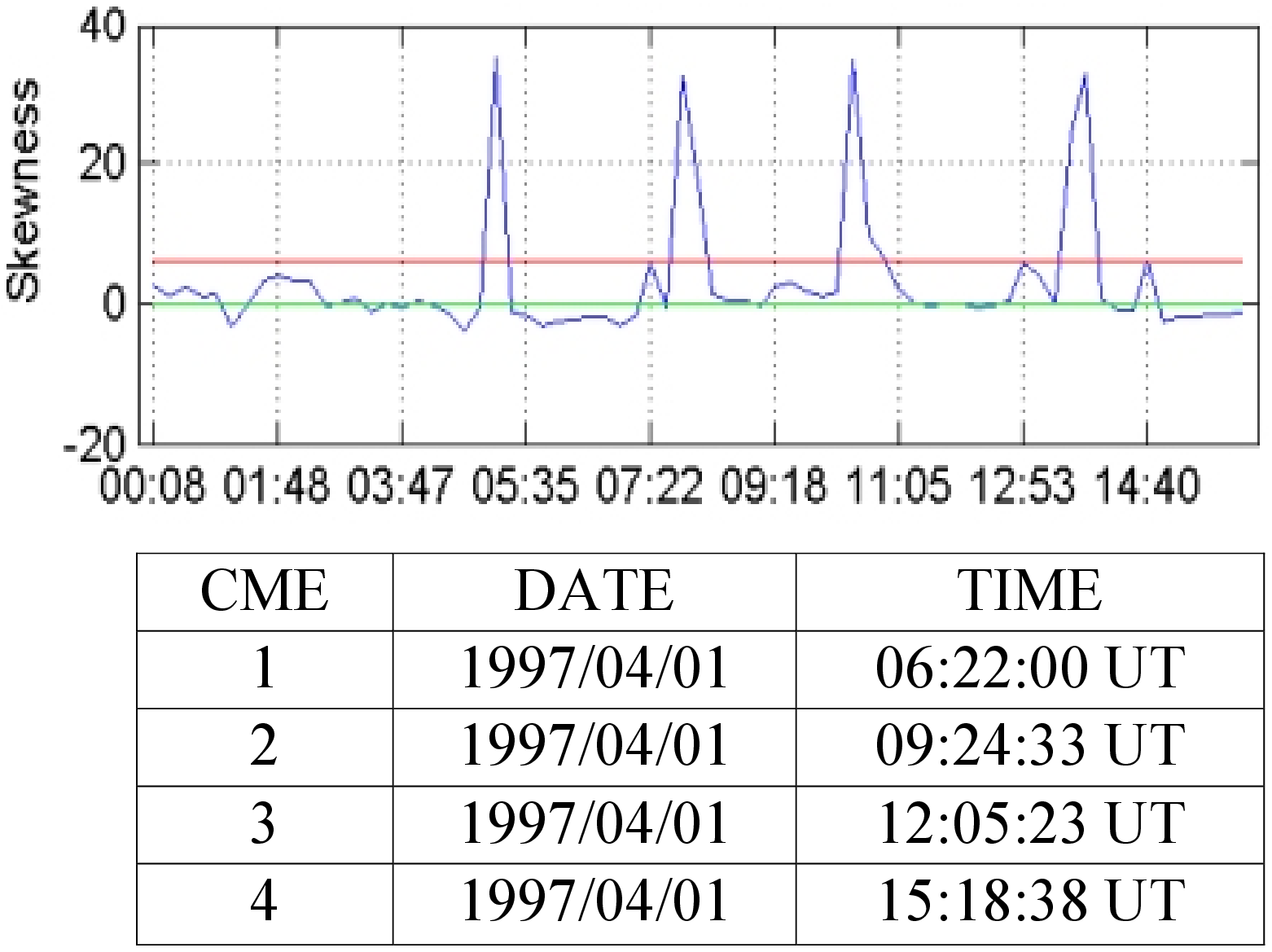}
               \hspace*{-0.014\textwidth}
              }
     \vspace{-0.37\textwidth}   
     \centerline{ \scriptsize   
      \hspace{0.04\textwidth}   {\bf \large (a)}
      \hspace{0.4 \textwidth}  {\bf \large (b)}
         \hfill}
     \vspace{0.33\textwidth}    
\caption{High cadence synoptic observations of solar disk in EIT/SOHO 195 $\dot{A}$ wavelength by SOHO. Every skewness burst corresponds to CME initiation, observed this day. {\bf (a)} - 195 $\dot{A}$  wavelength EIT image at 06:22~UT. {\bf (b)} - Time series of asymmetry (skewness) constructed for every image of the day 01 April, 1997. The time in the list corresponds to the first appearance of CME in the white light LASCO coronograph.}
   \label{Fig1}
   \end{figure}

NEMO algorithm consists of two main parts that are briefly presented below.

{{\bf{Event Detection:}} The initial part refers to the detection of an event in the EUV disc and it is based on the modifications of
basic statistical properties of the image sequences during such events. All the discovered type of CME precursors in EUV, such as
various EIT waves and the sudden loop opening events are triggered by solar flares or brusque filament disappearance and they
always contain eruptive dimmings. As the consequence, all these events have one important common characteristic: they strongly affect the probability distribution functions (PDF) of pixels distributions. The techniques that allowed the extraction of the eruptive features from the EUV solar disk under complex solar conditions are based on the analysis of skewness and kurtosis. Skewness is the measure of the asymmetry of the
probability distribution function (PDF). If the left tail of the PDF is more pronounced than the right one, then
the PDF has negative skewness and when the reverse is true, it has positive skewness. Kurtosis measures the
excess probability (flatness) in the tails, where excess is defined in relation to a Gaussian distribution.
Large values of these high order moments indicate the existence of intermittent/bursty events which
are characterized by strongly non-Gaussian PDFs \cite{Sandberg2009}. The sudden appearance of coherent and intensive structures on the Sun is reflected by bursts of skewness and kurtosis of the pixel distribution. The appearance of these bursts is a reliable indicator of a flare
or an eruption occurrence.}

{Solar Eruptions are barely seen in the image due to their low intensity contrast with respect to simultaneously appeared solar flare.
Thus, fixed differences (FD) images are constructed by the subtraction of a reference image of the day and taking into account the solar
differential rotation compensation.  The first step in NEMO consist of the detection of higher order moments bursts of differences images distribution. This detection principle is very robust and can be applied successfully in data of any EUV telescope especially for the case of
large scale events. If these moments grows in few consecutive images an event is detected.}

{As an example, let us consider \textit{``a worst case''} of the four faint CMEs observed during 1997 April, 1 by LASCO coronograph.
The recognition of their precursors is not distinctive by visual inspection on the EUV/SOHO 195 $\dot{A}$ images
of the full Sun (Fig.~\ref{Fig1}\textit{a}).
However, the asymmetry and the flatness of the pixel distributions on the differences images are excellent indicators of
their occurrence (Fig.~\ref{Fig1}\textit{b}). Remarkably, one can see that these CME signatures are observed a up to a couple of hours
prior to the associated coronograph observations.}

\vspace{1.2 mm}
{\bf{Eruptive Dimming Recognition:}} Once the occurrence of an event has been detected,
the next step is to recognize if an {\it eruptive dimming} is present on the
FD image of solar disk. Eruptive dimmings appear as regions of dark intensity in EUV wavelengths. However, there are many
regions of dark intensity that are present continuously on the EUV Sun. The eruptive dimmings differs from them as they appear
rapidly and they have the tendency to expand much faster.

\begin{figure}    
   \centerline{\hspace*{0.01\textwidth}
               \includegraphics[width=0.85\textwidth,clip=]{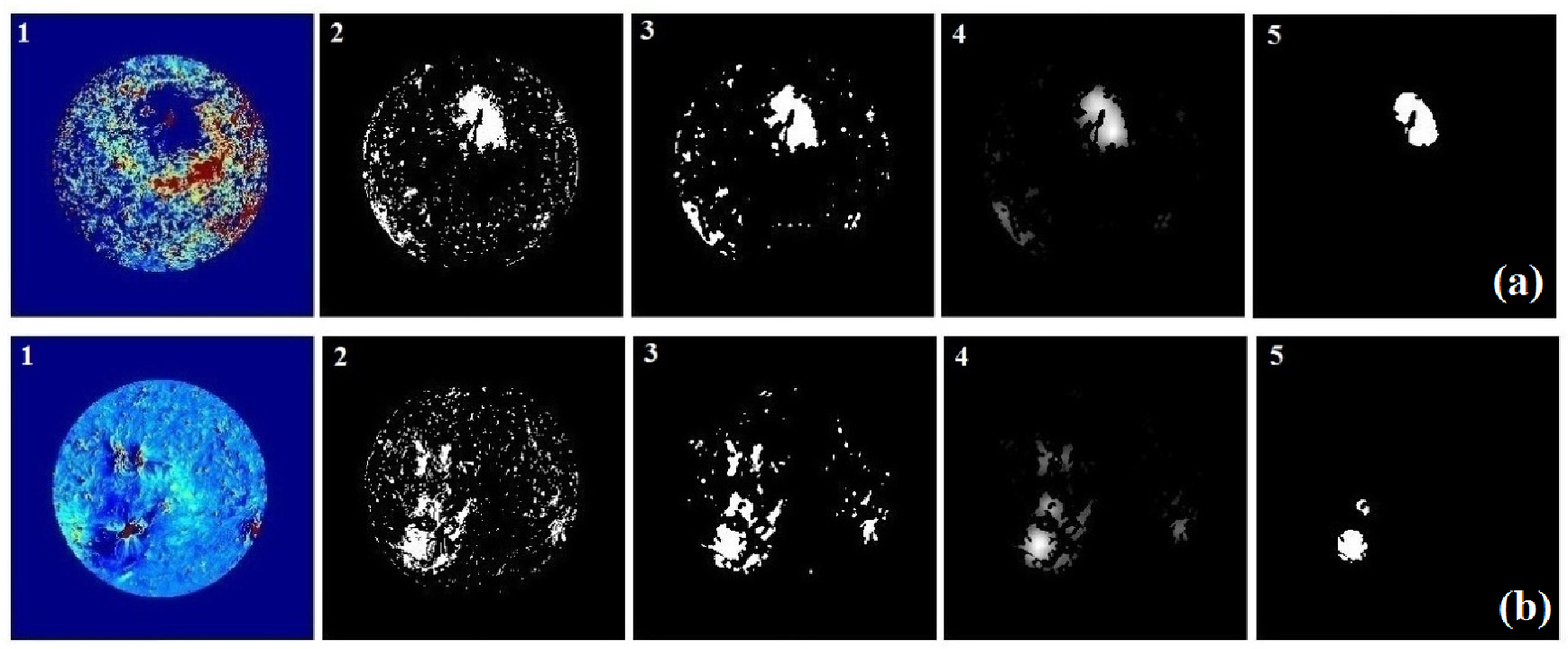}
               \hspace*{-0.014\textwidth}
              }
\caption{ 5 steps of dimming extraction form EUV/SOHO 195 $\dot{A}$  EUV corona image. {\bf (a)} -- 12 May, 1997 at 05:07~UT; {\bf (b)} -- 07 April, 1997 at 14:22~UT.}
   \label{Fig2}
   \end{figure}

The algorithm of eruptive dimming recognition includes the following consecutive steps:
\vspace{1.5 mm}
\begin{enumerate}
\item [{\it 1.}]
{\it The construction of fixed differences,} with solar differential derotation compensation included, where
 the first image before the solar event is subtracted from the sequence of the subsequent images.
 \vspace{1.2 mm}

 \item [{\it 2.}]
{\it The extraction of regions with decreased intensity} from the difference images: This region is extracted by selecting the 5\%
of the darkest pixels in the FD image. The resulted images contain large-scale distinctive regions of reduced intensity.
However, many small- scale scattered regions of reduced intensity  - attributed to the presence of noise
effects - still remain.
 \vspace{1.2 mm}

  \item [{\it 3.}]
{\it The median filtering application} for the reduction of small scale noise: Using this standard procedure the small
scale scattered noisy points are further reduced.
 \vspace{1.2 mm}

   \item [{\it 4.}]
{\it Clustering} of decreased intensity regions: Dimmings are the large scale connected regions of reduced intensity with
respect to the scattered points attributed to noise. For the extraction of the dimming region the agglomerative filtering is applied; for each
decreased intensity reference pixel with coordinates $x_{i} ,y_{i}$ the agglomerative weight $W(x_{i} ,y_{i} )$ is computed.
This weight is equal to the number of pixels with reduced intensity in the limited square vicinity around the reference one
and permits the separation between the small scattered and the large scale regions of negative intensity.

\vspace{0.2 mm}

  \item [{\it 5.}]
{\it Dimming Extraction.}\\ To extract the larger dimming area from the background of decreased intensity regions, we set the maximum weight  $M=\mathop{\max }\limits_{i} $$\{ W(x_{i} ,y_{i} )\} $ and we select those pixels with weight that exceeds $q*M$, where $q\in [0;1]$ is the coefficient of agglomeration.\\

Figure \ref{Fig2} shows all the five steps of dimming extraction as described above using EIT/SOHO 195~$\dot{A}$ images taken at 1997 May, 12
at 05:07~UT (top panel) and 1997 April, 7 at 14:22~UT (bottom panel). If the resulted filtered image contains more then one
dimming area, the largest one is selected for the subsequent analysis.
The decision whether a dimming is eruptive or not is received depending on the growth of its dimming area
in few images.

\vspace{1.5 mm}
\end{enumerate}

\begin{description}
\item
The algorithm of eruptive dimming extraction can be summarized as follows:
\vspace{1.2 mm}
\begin{itemize}
\item
Positive answer on the first part indicated the detection of burst in moments of pixel distribution behavior and
provides an alert about Solar Flare occurrence.
\vspace{1.2 mm}
\item
Positive answer on the second part indicates a growth in the surface area of the dimming and
provides alert about Solar Eruption occurrence and prompt appearance of CME in the Heliosphere.
\end{itemize}
\end{description}
It is evident that the identification of dimming regions may be the earliest and most efficient method for the prediction of CMEs.

\section{The modification of dimming extraction algorithm}

The dimming extraction by NEMO as presented above was based so far on the estimation of dimming surface in terms of the
numbers of pixels. However,
areas of equal surface at the image center and at the solar limb may differ up to five times in terms of pixel numbers
due to projection effects. As a consequence, this may lead to unjustified comparison of regions with decreased intensity and finally to the loss of
events occurred near the limb in favor of smaller dimming located in the center of the disc.

The clustering technique (agglomerative filtering) of decreased intensity regions uses the square vicinity around a center pixel and the information is taken from an area which is actually the projection of a square to a hemisphere. Thus, it can significantly distort the filtering result in the resulting area forms due to the asymmetry of the vicinity relatively to its center.

The extraction of eruptive dimming is carried out by comparing the size of the decreased intensity regions but not the intensity itself. However, the intensity of the dimming is also a significant characteristic as the area \cite{Bewsher2008}. The released energy during the dimming formation is almost equal to the energy of CME \cite{Fillipovbook2007}. The joint consideration of dimming areas and intensities can provides an estimation on the power of CME \cite{Zhukov2004b} and therefore to increase the reliability of the eruptive dimming extraction.

\vspace{1.5 mm}

In order to improve the detection efficiency of NEMO, the following updates have been applied:

\begin{itemize}
\vspace{1.2 mm}
\item
{\it Estimation of dimming areas in $km^2$. } The dimming area is computed now directly in terms of square kilometers instead of the EIT (or EUVI) pixels by taking into account that an EUV image is a projection of the solar sphere. Using the surface integral and the mean value theorem, we calculate the surface of $i$ pixel in $km^{2} $ using the following relation:
\[S_{i} =\frac{R^{2} }{r\sqrt{r^{2} -x_{i}^{2} -y_{i}^{2} } }\]
where $R$ is solar radius in $km^{2} $ and $r$ is  the solar radius in pixels. The total dimming area in $km^{2}$ is simply determined by the
sum of the pixels surface that form the dimming area.

\vspace{1.2 mm}

\item
{\it Clustering in circle vicinity.} The clustering procedure is significantly improved since the agglomerative filtering is carried out
on the spherical surface of the circular vicinity of the square vicinity. We have found that the circle vicinity clustering
improved drastically the estimation of the dimming structures with the maximum possible accuracy.
\vspace{1.2 mm}

\item
{\it Complex characteristic for dimming extraction} connected to CMEs is modified. The NEMO algorithm uses a criterion based on the surface
of the dimming area for the extraction of the dimming. In order to increase the confidence on dimming extraction, we have
introduced a new complex characteristic which allows the simultaneous
consideration of both area and intensity of dimmings. This characteristic can be considered as
a ``volume metric'' of the dimming and it is defined by the following relation:

\[V=\sum _{D}I_{i} \cdot  S_{i} [km^{2} \cdot cnt\cdot pxl^{-1} ]\]

Here $D$ is the surface of the dimming and $I_i$ is the intensity of the dimming pixels.
This characteristic variable can be used for the construction of more qualitative picture about the
dimming evolution in time and can provide an rough estimation on the CME power which is connected with
the formation of considered dimmings.

\end{itemize}

\section{Comparative analysis between the tested and basic algorithms}

A comparative analysis between the updated and the basic NEMO algorithms has been carried out based
on a number of SOHO and STEREO EUV observations of solar corona.
It was clearly shown that the new algorithm provides a more accurate estimation of the shape and
the size of eruptive dimmings for all the considered cases. The modified algorithm is more sensitive
to small-scale events and it provides much earlier detection while the basic algorithm could miss
completely small scale events or detect larger ones with additional delays.
Remarkably, the results are drastically improved especially for the events observed near the solar limb.
Three characteristic examples are presented below.

\begin{figure}    
   \centerline{\hspace*{0.01\textwidth}
               \includegraphics[width=0.20\textwidth, angle=270, clip=]{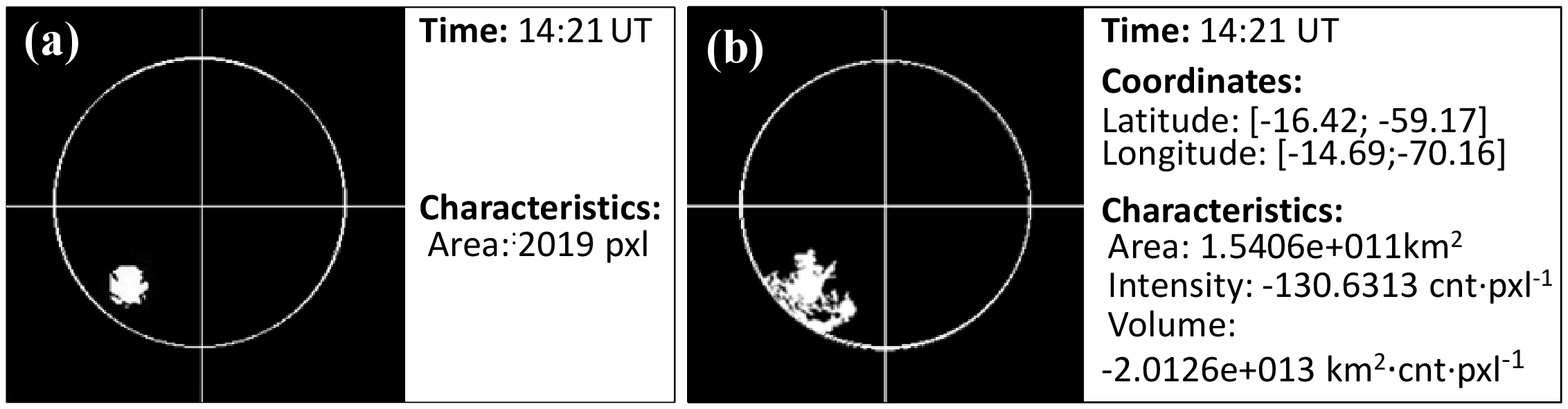}
               \hspace*{-0.014\textwidth}
              }
\caption{More precise extraction of the dimming of 7 April, 1997 EIT/SOHO event by modified algorithm with respect to the dimming extracted by the basic NEMO method {\bf (a)}. {\bf (b)} -- More dimming characteristics are provided by modified extraction algorithm.}
   \label{Fig3}
   \end{figure}

\vspace{1.2 mm}
{\bf Improved processing of dimmings:} An eruptive event was observed near the solar limb by EIT/SOHO imager in 195~$\dot{A}$ wavelength 7 April, 1997 at 14:22~UT. This event was automatically detected in both basic and modified algorithms. Fig.~\ref{Fig3} depicts the extracted structures from the differences images using the basic (Fig.~\ref{Fig3}\textit{a}) and the modified algorithm (Fig.~\ref{Fig3}\textit{b}). As it is shown, the modified algorithm demonstrates considerably more precise definition of the dimming shape.
The surface of the dimming is given in $km^2$ instead of pixels while a number of additional parameters is also provided.

\begin{figure}    
   \centerline{\hspace*{0.01\textwidth}
               \includegraphics[width=0.80\textwidth,clip=]{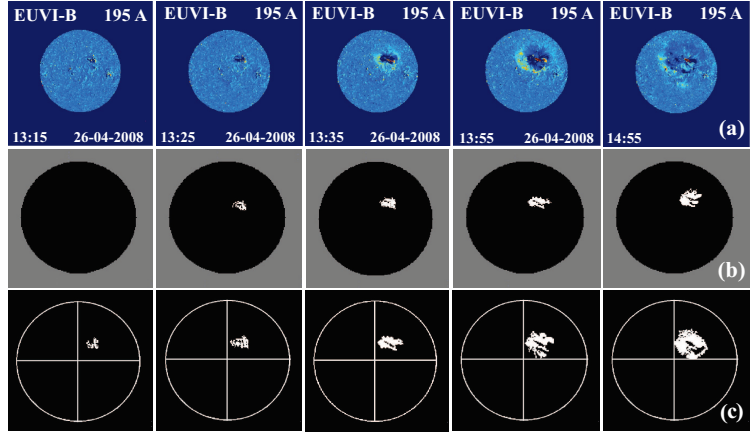}
               \hspace*{-0.014\textwidth}
              }
\caption{26 April, 2008 EUVI/STEREO event, detected with the delay by basic algorithm.  {\bf (a)} -- Differences images.  {\bf (b)} -- Dimmings extracted by basic algorithms. {\bf (c)} -- Dimmings extracted by modified algorithm with characteristics indicated in the Table~\ref{table1}.}
   \label{Fig4}
   \end{figure}

\vspace{1.2 mm}
{\bf Improved detection efficiency:} The EUVI/STEREO-B observations of solar disk eruptive event that occurred 2008 April, 26 from 13:15~UT to 13:55~UT are shown in Fig.~\ref{Fig4}. The top panel depicts the fixed differences images of the event created by subtracting a fixed reference image before the event onset from the subsequent images. The middle and bottom panels show the dimmings as extracted by using the basic and the modified algorithms respectively. As it is shown in Fig.~\ref{Fig4}, the modified version of the algorithm detects the dimming from the early stages
of the event development (at 13:15~UT) in contrast to the basic algorithm which provides detection alert with a relative delay of 10 minutes. The parameters of the event are shown in Table~\ref{table1} as extracted by the application of the new algorithms.
The basic algorithm can not provide in such accuracy the shape and the rest characteristic parameters of the eruptive dimming.

\begin{figure}    
   \centerline{\hspace*{0.01\textwidth}
               \includegraphics[width=0.80\textwidth, clip=]{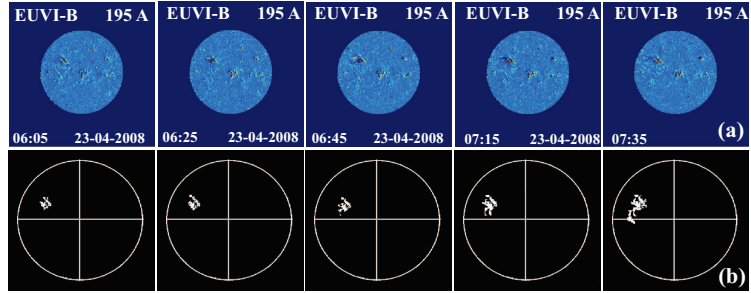}
               \hspace*{-0.014\textwidth}
              }

\caption{ Small-scale dimming of SECCHI/STEREO event,  detectable only by modified algorithm. {\bf (a)} --  06:05~UT--06:35~UT differences images of 23 April, 2008 EUVI/STEREO-B event. {\bf (b)} -- Extracted dimmings with the characteristics given in the Table~\ref{table2}.}
   \label{Fig5}
   \end{figure}

%
%
\begin{table}
\begin{center}
\caption{Characteristics of 26 April, 2008 EUVI/STEREO event {\it eruptive dimming} extracted by modified NEMO algorithm.}
\begin{tabular}{|c|c|c|c|c|c|} \hline
     \textbf{Time}  & Latitudes & Longitudes & Area & Intensity & Volume \\ \hline

13:15 & 9.09; 20.46 & 11.86; 22.25 &  8.86e+009 & -1.48e+002  &  -1.31e+012  \\
13:25 & 9.23; 20.48 & 3.08; 21.71 &  1.09e+010 & -2.70e+002  &  -2.95e+012  \\
13:35 & 10.26; 23.41 & 3.07; 21.15 &  3.69e+010 & -3.78e+002  &  -6.40e+012  \\
13:55 & 1.63; 28.55 & 0.78; 27.76&  5.14e+010 & -5.37e+002  &  -2.76e+013 \\
14:05 & 0.89; 27.76 & 0.15; 30.68 &  6.75e+010 & -4.82e+002  &  -3.25e+013  \\    \hline
{\scriptsize UT}    &  {\scriptsize degrees}  & {\scriptsize degrees} & {\scriptsize $km^2$} &  {\scriptsize $cnt \cdot pxl^{-1}$}  &  {\scriptsize $km^2 \cdot cnt \cdot pxl^{-1}$}\\ \hline

\end{tabular} \label{table1}
\end{center}
\end{table}
%
\vspace{1.2 mm}
{\bf  Detection of small-scale events:} {The improved spatial and temporal resolution of the EUVI telescope on-board of STEREO spacecrafts has permitted the observation of smaller eruptive events with respect to those observed before the upgrade of the instrument \cite{Innes2009,Podladchikova2010,Innes2010}. A relatively small-scale event was captured by the EUVI/STEREO-B telescope on 2003 April, 23 from 06:05~UT to 07:35~UT. The basic algorithm could not issue an alert of eruptive dimming for this case while the updated algorithms did it with success.
The fixed differences and the extracted dimming by the modified algorithm are shown on the  top and bottom panel of
Fig.~\ref{Fig5} respectively, while the characteristics of the extracted event are given in Table~\ref{table2}.}

%
%
\begin{table}
\begin{center}
\caption{Extraction Characteristics of 23 April, 2008 small-scale EUVI/STEREO event, detectable only by modified algorithm.}
\begin{tabular}{|c|c|c|c|c|c|} \hline
     \textbf{Time}  & Latitudes & Longitudes & Area & Intensity & Volume \\ \hline

06:05 & 9.33; 21.40  & -31.39; -41.48  & 9.38e+009  & -1.79e+002  & -1.68e+012  \\
06:25 & 9.31; 23.87 & -29.38; -41.49  & 1.22e+010  & -3.72e+002 & -4.52e+012   \\
06:45 & 2.32; 23.91 & -28.35; -44.57 & 1.67e+010  & -4.41e+002 & -7.36e+012  \\
07:15 & 1.55; 23.93 & -26.36; -45.95  & 2.06e+010 & -5.46e+002  & -1.13e+013 \\
07:35 & -3.87; 23.95 &  -26.20; -48.03 & 2.08e+010  & -5.79e+002  & -1.20e+013   \\    \hline
{\scriptsize UT}    &  {\scriptsize degrees}  & {\scriptsize degrees} & {\scriptsize $km^2$} &  {\scriptsize $cnt \cdot pxl^{-1}$}  &  {\scriptsize $km^2 \cdot cnt \cdot pxl^{-1}$}\\ \hline

\end{tabular} \label{table2}
\end{center}
\end{table}
%

{From all the examples presented above it becomes evident that the consideration of both surface of the dimming region
and dimming intensity in combination with the more accurate extraction method  of circle vicinity
clustering in spherical coordinates increase significantly the detection efficiency and the information
provided by the extracted dimmings. An updated version of NEMO that will include all the algorithms presented
is expected to introduce the following additional and more accurate information in the NEMO catalogs of eruptive events:

\vspace{1.2 mm}
\begin{itemize}
\item
Dimming coordinates (latitudes and longitudes of the dimming borders)
\item
Dimming intensity
\item
"Volume metric" of dimming
\item
Surface of the dimming area in km$^2$
\end{itemize}
\vspace{1.2 mm}
The dimming coordinates listed in NEMO catalogs will provide systematic information about the CME
precursors while the other parameters of dimmings: area, intensity, "volume metric" may provide
essential information for the estimation of CME power.}

\section{Conclusions}

We have developed novel algorithms for the NEMO detection tool of flares and CMEs for
post-processes of EIT/SOHO and SECCHI/STEREO EUV solar disk images. NEMO version that is currently in
operation consists of two main steps, namely the Event Detection and the Eruptive Dimming Recognition.
The event detection is based on the detection of bursts in the high order moments of the image while the
eruptive dimming extraction is based on a sequence of filtering techniques applied in the pixel distributions
of image sequences. In this work, we have presented a series of updates in the eruptive dimming extraction algorithms
that allow us to increase significantly the detection efficiency of eruptive dimmings linked to CMEs.

The surface of the dimming area is computed now directly in terms of physical variables (square kilometers) by taking
rigorously into account that EIT images correspond to solar sphere projections. The clustering of dark regions
is achieved through circle vicinity clustering on latitude and longitude. Furthermore, the novel methods for
the eruptive dimming extraction - based on the volume metric of the dimming increase the detection efficiency
and the accuracy of the associated extracted parameters.
Using a series of examples, we have shown that the modified version of NEMO tool presents indeed significantly higher temporal and
spatial efficiency on the automatic detection of CME precursors. In particular, small eruptive events located near the solar limb can
be detected now while major events can be detected at earlier times than before.
The NEMO tool will incorporate the optimized new algorithms and it is expected to provide early warnings for CMEs precursors
and automatically construct new catalogs with enhanced and more accurate information about the detected solar eruptive events.

\begin{acks}
A.V. and P.L. acknowledge the ASBL--VZW research grant for MSc studies on the "Dynamics of the Solar System" at the Royal Observatory of Belgium.
\end{acks}

%
%

\end{article}
\end{document}